\documentstyle[prl,twocolumn,aps,psfig]{revtex}
\input{epsf}

\newcommand{\ez}{\vec{e}_z}
\newcommand{\vu}{\vec{u}}
\newcommand{\beq}{\begin{equation}}
\newcommand{\eeqn}[1]{\label{#1}\end{equation}}
\newcommand{\dt}[1]{\frac{\partial  #1}{\partial t}}
\newcommand{\na}{ \vec{\nabla} }
\newcommand{\noi}{ \noindent }
\newcommand{\eq}[1]{(\ref{#1})}
\newcommand{\ie}{{\it i.e.}\ }
\def\bbbn{{\rm I\!N}} 
\newcommand{\tv}{ \rightarrow}
\newcommand{\lp}{ \left(}
\newcommand{\rp}{ \right)}

\begin{document}

\draft
\twocolumn[\hsize\textwidth\columnwidth\hsize\csname @twocolumnfalse\endcsname

\title{Waves attractors in rotating fluids: a paradigm for ill-posed
Cauchy problems}

\author{M.~Rieutord$^{1,2}$, B.~Georgeot$^3$ and L.~Valdettaro$^4$}

\address {$^1$Observatoire Midi-Pyr\'en\'ees, 14 avenue E. Belin, F-31400
Toulouse, France \\
$^2$Institut Universitaire de France\\
$^3$Laboratoire de Physique Quantique, UMR 5626 du CNRS, 
Universit\'e Paul Sabatier, F-31062 Toulouse Cedex 4, France\\
$^4$Dip. di Matematica, Politecnico di Milano, Piazza L. da Vinci, 32, 20133 
Milano, Italy}

\date{\today}

\maketitle

\begin{abstract}
In the limit of low viscosity, we show that the amplitude of the modes
of oscillation of a rotating fluid, namely inertial modes, concentrate
along an attractor formed by a periodic orbit of characteristics of
the underlying hyperbolic Poincar\'e equation.
The dynamics of characteristics is used to elaborate a
scenario for the asymptotic behaviour of the eigenmodes and
eigenspectrum in the physically relevant r\'egime of very low
viscosities which are out of reach numerically. This problem offers a
canonical ill-posed Cauchy problem which has applications in other
fields.
\end{abstract}

\pacs{PACS numbers: 47.32.-y, 05.45.-a, 02.60.Lj, 04.20.Gz}
\vskip1pc]

Rotating fluids encompass all fluids whose motions are dominated by
the Coriolis force. These flows play an important role in astrophysics
or geophysics where the large size of the bodies makes the Coriolis
force a prominent force. Some engineering problems like the
stability of artificial satellites also require the study of rotating fluids
because of their liquid-filled tanks \cite{Mana92}. This latter problem
is related to the existence of waves specific to rotating fluids,
namely inertial waves, which easily resonate. These waves play also
an important part in the oscillation properties of large bodies like
the atmosphere, the oceans, the liquid core of the Earth\cite{rieu99},
rapidly rotating stars\cite{DR00} or neutron stars\cite{AKS99}. As
such, they have been considered since the work of Poincar\'e on the
stability of figures of equilibrium of rotating masses
\cite{Poinc1885}. Pressure perturbations of inertial modes for inviscid
fluids obey the Poincar\'e equation (PE) (christened by
Cartan\cite{cartan22}) which reads  $\Delta P -(2\Omega/\omega)^{-2}
\partial^{2}P/\partial z^{2}=0$ where $\Omega\ez$ is the angular
velocity of the fluid and $\omega$ is the frequency of the oscillation.
Since $\omega<2\Omega$ \cite{Green69}, the PE is hyperbolic (energy
propagates along characteristics) and since its solutions must meet
boundary conditions, the problem is ill-posed mathematically. Although
some smooth solutions exist (for instance for a fluid contained in a
full sphere or a cylinder), one should expect singular solutions in the
general case. These latter solutions have been made explicit only
recently thanks to numerical simulations which include viscosity to
regularize the singularities and let this parameter be very small as in
real systems\cite{RV97,RGV99}.

In this letter we wish to present a scenario, based on analytical and
numerical results, for the asymptotic behaviour of inertial modes at
small viscosities. We use the case of a spherical shell as a container,
which is relevant for astrophysical or geophysical problems, but it
will be clear that this case is general.  We will only sketch the main
results, more details can be found in \cite{RGV99}.  While the fluid
mechanical problem is of much interest by itself, it opens new
perspectives in the theory of Partial Differential Equations (PDE) and
also offers a toy model for some (very involved) problems of General
Relativity which we shall present briefly.


The model we use is a spherical shell whose inner radius is $\eta R$ and outer
radius $R$ ($\eta <1$). The fluid is assumed incompressible with a
kinematic viscosity $\nu$. We write the linearized equations of motion for small
amplitude perturbations for the velocity $\vu$
in a frame corotating with the fluid; momentum
and mass conservation imply:

\beq \dt{\vu} + \ez\times\vu = -\na p + E\Delta\vu, \qquad \na\cdot\vu
=0\eeqn{master}

\begin{figure}
\epsfxsize=2.5in
\epsfysize=2.5in
\epsffile{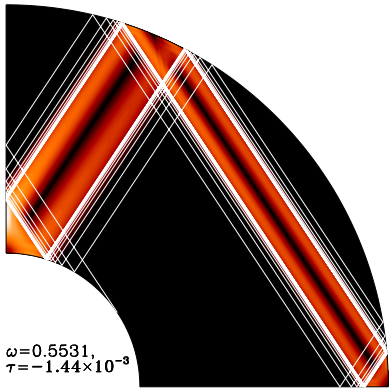}
\caption[]{Kinetic energy in a meridional section of a spherical shell
of an inertial mode in a viscous fluid. For this numerical solution,
$E=10^{-8}$, 570 spherical harmonics and 250 Chebyshev polynomials have
been used (the numerical method is described in \cite{RV97}). The mode is 
axisymmetric and symmetric with respect to
equator. $\eta=0.35$ like in the Earth's core. $\omega$ is the frequency
of this mode and $\tau$ its damping rate. Stress-free boundary
conditions are used. The convergence of characteristics towards the
attractor is also shown (white lines).}
\label{fig1}
\end{figure}

\noi when dimensionless variables are used; $(2\Omega)^{-1}$ is the time
scale and $E=\nu/2\Omega
R^2$ the Ekman number. When $E$ is set to zero and $\vu$ is eliminated,
one obtains the Poincar\'e equation. In nature $E\ll1$ and one is
tempted to use boundary layer theory and singular perturbations to solve
\eq{master}. However, this is feasible only when regular solutions exist
for $E=0$; this is the case when the container is a full
sphere\cite{Green69} but not when the container is a spherical shell.
Indeed, numerical solutions of the eigenvalue problem issued from
\eq{master}, where solutions of the form $\vu(\vec{r})e^{\lambda t}$ are
searched for (with $-1\leq \omega=Im(\lambda)\leq 1$), yield eigenmodes
of the kind shown in Fig.~\ref{fig1}.  In this figure we see that the
amplitude of the mode is all concentrated along a periodic orbit of
characteristics of the PE; we found this property to be quite general, after
extensive numerical exploration of least-damped modes of \eq{master}
\cite{RV97,RGV99},
and will now explain its origin and consequences on the asymptotic
spectrum of inertial modes. For this purpose we will use axisymmetric
modes since the azimuthal dependence of solutions can always be
separated out because of the axial symmetry of the problem.

\begin{figure}
\vspace*{5mm}
\epsfxsize=3.5in
\epsfysize=2.5in
\centerline{\epsffile{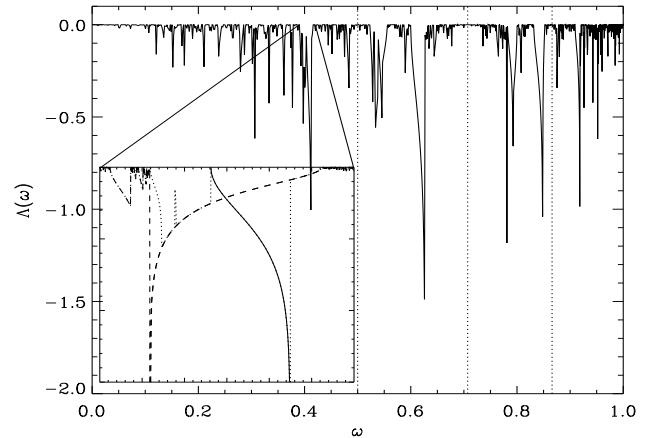}}
\caption{LE $\Lambda(\omega)$ of the orbits as a function of $\omega$
for $\eta=0.35$.
Inset: blow-up showing the LE of two coexisting attractors (full and
dashed thick lines). }
\label{lyap}
\end{figure}

For understanding the concentration of kinetic energy along a periodic
orbit of characteristics, it is necessary to consider in some details
the dynamics of these lines. Characteristics of PE are, in a meridional
plane, straight lines making the angle $\arcsin\omega$ with
the rotation axis. A numerical calculation of their trajectories shows
that they generally converge towards a periodic orbit which we call,
after\cite{ML95}, an {\em attractor}. The periodic orbit of
Fig.~\ref{fig1} is one example of such an attractor.
The Lyapunov exponent (LE) of a trajectory, defined by $ \Lambda =
\lim_{N\rightarrow\infty} \frac{1}{N}\sum_{n=1}^N
\ln\left|\frac{d\phi_{n+1}}{d\phi_n}\right| $ ($\phi_n$ is the latitude
of the n$^{\rm th}$ reflection point), describes how fast characteristics
are attracted or repelled.  Its computation as a function of
frequency shows that attractors ($\Lambda < 0$) are ubiquitous in
frequency space (see Fig.~\ref{lyap}).  Their existence shows that the
dynamical system described by the characteristics is not hamiltonian;
the ``dissipation" is purely geometrical and is due to the fact that,
unlike billiards, the reflection on boundaries is not specular but
conserves the angle with the rotation axis. In fact, the dynamics of
rays is a one-to-one one-dimensional map (from the outer boundary to
itself), piecewise smooth, but with a finite number (twelve) of
discontinuities.  This kind of map has not been studied in the
literature of dynamical systems, perhaps because it does not produce
chaos because of its invertibility.  Iterations of such a map generate
fixed points which either correspond to attractors or to
some neutral periodic orbits.
Indeed, if $\eta=0$ (\ie the sphere is full), all orbits
such that $\omega=\sin(p\pi/2q)$ with $(p,q)\in \bbbn^2$, are neutral
and periodic while those such that $\omega=\sin(r\pi)$, $r$ being
irrational, are neutral ergodic (quasiperiodic). When $\eta$ is
non-zero only a finite number of such neutral periodic orbits subsist; for
instance, if $\eta=0.35$ which is the aspect ratio of the Earth's
liquid core, $q=1,2,3,4$ are the only possibilities.  Interestingly, we
face here a situation which is just the opposite of the one described
by the KAM theorem in Hamiltonian systems:  when the full sphere is
perturbed by the introduction of an inner sphere, all ergodic orbits
are instantaneously destroyed while the longer periodic orbits survive
the smaller the denominator $q$ is.

Apart from these isolated frequencies which become rarer and rarer as $\eta$
increases, generic trajectories are in the basin of attraction of
attractors. We were able to show \cite{RGV99} that the number of attractors at a
given frequency is finite. The inset of Fig. \ref{lyap} shows the typical case
where an attractor exists in a frequency band $[\omega_1,\omega_2]$
with $\Lambda(\omega_1)=0$, $\Lambda(\omega_2)=-\infty$ and
$\omega_2-\omega_1\sim 1/N^2$ where $N$ is the length of the attractor
defined as its number of reflection points.
Near $\omega_1$, $\Lambda\sim\sqrt{\omega-\omega_1}$ and near
$\omega_2$, $\Lambda(\omega)\sim {1\over N}\ln N(\omega-\omega_2)$. The
latter implies that long attractors have small LE in a large fraction
of $[\omega_1,\omega_2]$ (all these results are shown in
\cite{RGV99}).

The existence of attractors for characteristics implies that solutions of
the inviscid problem (\ie of PE) are singular. This
property can be made explicit in the simplified case of a 2D problem.
Indeed, in this case the PE may be written
$\partial^2P/\partial u_+\partial u_-=0$ using characteristics
coordinates; solutions may be constructed explicitly from an  arbitrary
function but, as shown in \cite{Schaef75}, regular eigenmodes exist
only when neutral periodic orbits exist and eigenvalues are infinitely
degenerate. When attractors are present, the scale of variations of the
pressure vanishes on the attractors while its amplitude remains
constant. As velocity depends on the pressure gradient, it diverges on
the attractor; this divergence is like the inverse of the distance to the
attractor which makes the velocity field not square integrable. This
result seems to be valid also in 3D \cite{RGV99}.

We therefore understand why solutions of \eq{master} look like Fig.~1: the
inviscid part of the operator focuses energy of the modes  thanks to the
action of the mapping made by characteristics while viscosity opposes to
this action via diffusion. The resulting picture of Fig.~1 therefore
comes from a balance between inviscid terms and viscous ones; let us
make this more quantitative.

\begin{figure}
\epsfxsize=3.5in
\epsfysize=1.5in
\centerline{\epsffile{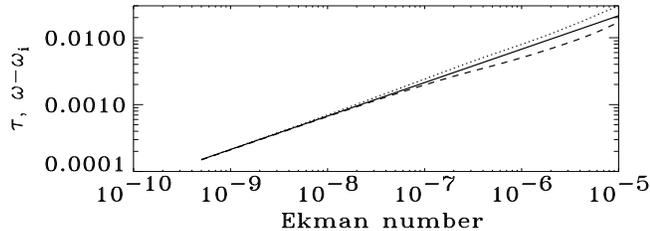}}
\caption{Asymptotic behavior of an eigenvalue. The dashed line is
$\omega-\omega_i$ as a function of $E$, while the dotted line is for the
damping rate $\tau$. The solid line represents the `theoretical' law
$E^{1/2}$. $\omega_i=0.403112887$ is a root of $\Lambda(\omega)=0$ when
$\eta=0.35$.}
\label{fig5}
\end{figure}

For this purpose we first observe that the patterns drawn by the kinetic
energy of the mode in Fig.~\ref{fig1} is in fact a shear layer whose
width scale with $E^\sigma$ and $\sigma\simeq 1/4$. Such a scaling
is observed numerically and seems generic \cite{RV97,RGV99}; it implies
that the damping rate of such modes scales like $E^{1/2}$ as clearly
shown in Fig.~\ref{fig5}.  Now we may consider a wave packet
travelling around an attractor in a slightly viscous fluid. The above
mentioned balance, when applied to both the width and the amplitude of
the packet, leads to a relation between the LE and the Ekman number
such as $\Lambda \sim E^{1-3\sigma}$ with $\sigma < 1/3$ for an
eigenmode of the viscous problem. We see that the constraint $\sigma <
1/3$ is met by actual shear layers. It therefore turns out that
frequencies of eigenmodes of the viscous problem are such that
$\Lambda\tv0$ when $E\tv0$ which means that they will gather around the
roots of the equation $\Lambda(\omega)=0$.

The above result shows the importance of the scaling verified by shear
layers. A boundary layer analysis reveals that these shear layers are in
fact nested layers which consist of an inner $\sigma=1/3$-layer
surrounded by a thicker layer. The inner 1/3-layer can be fully
explicited. Using coordinates along the shear layers ($x$) and
perpendicular to it ($y$), we find that the $\varphi$-component of
the velocity verifies $\frac{\partial^3 {u_\varphi}}{{\partial Y^3}} =
-i\frac{\partial  {u_\varphi}} {{\partial q}}$, with $Y=y/E^{1/3}$ and $q
= x/\sqrt{1-\omega^2}$ which is also the equation verified by the stream
function in a steady shear layer of a rotating fluid\cite{MS69}. Solutions
which vanish in $Y=\pm\infty$ are self-similar and of the form
$u_\varphi = q^\alpha H_\alpha\lp Y/q^{1/3}\rp$ with $H_\alpha(t) = \int_0^\infty
e^{-ipt}e^{-p^3}p^{-3\alpha-1} dp$.  Besides, $\alpha=-\frac{1}{3}$ is the
only admissible value to ensure a coherent evolution of the width and
amplitudes after reflection on a boundary.


We are now in position to propose a scenario for the asymptotic behaviour of
inertial modes when the viscosity vanishes. Eigenfunctions reduce to nested shear
layers concentrated along attractors while the associated eigenvalues converge
toward the frequency $\omega_i$ such that $\Lambda(\omega_i)=0$ for the associated
attractor.  Furthermore, we can constrain this convergence of
eigenfrequencies; indeed, since $\Lambda \sim \sqrt{\omega-\omega_i}$,
one finds that $\omega = \omega_i + aE^{2-6\sigma} + \cdots$ and
$\tau=Re(\lambda)=-bE^{1-2\sigma}$ when $E
\rightarrow 0$; Fig.~\ref{fig5} shows that this law agrees well with
the numerical results, in the case shown, with $\sigma=1/4$.

\begin{figure}
\centerline{\psfig{figure=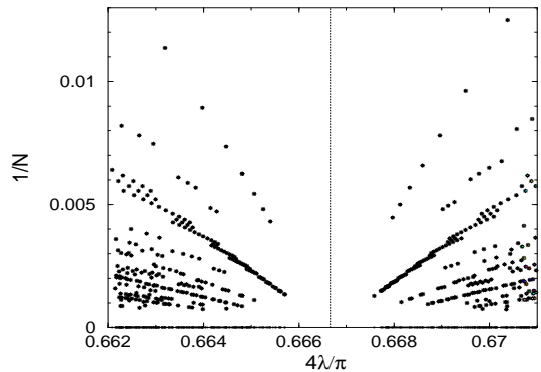,width=7cm,height=5cm}}
\caption{Inverse of the length $N$ of attractors with $N < 100$
for $\eta=0.35$, near the accumulation point $\pi/6$; each point
corresponds to an attractor with $\Lambda=0$ and therefore to a
point in the asymptotic spectrum.
Note the lengthening of the attractors as $\pi/6$ is approached.
Here $\eta=0.35$.} \label{length_att}
\end{figure}

In addition, we noticed earlier that for a finite number of $\omega$ such that
$\omega=\sin(p\pi/2q)$ all orbits of characteristics are periodic; this
implies that in the vicinity of these frequencies very long attractors
with very small average LE accumulate as shown by
Fig.~\ref{length_att}; therefore, these frequencies will be
accumulation points of the asymptotic spectrum. Moreover, around these
frequencies eigenmodes are weakly damped. On the contrary, modes whose
frequency is in the frequency band of short attractors (like the
one of Fig.~\ref{fig1}) are more strongly damped. It therefore turns out
that the LE curve in Fig.~\ref{lyap} will strongly constrain the
distribution of least-damped modes in the complex plane at finite
viscosities: such modes will avoid the large frequency bands of short-period
attractors and concentrate around frequencies where $\Lambda(\omega)=0$
especially those with $\omega=\sin(p\pi/2q)$.

\begin{figure}
\centerline{\psfig{figure=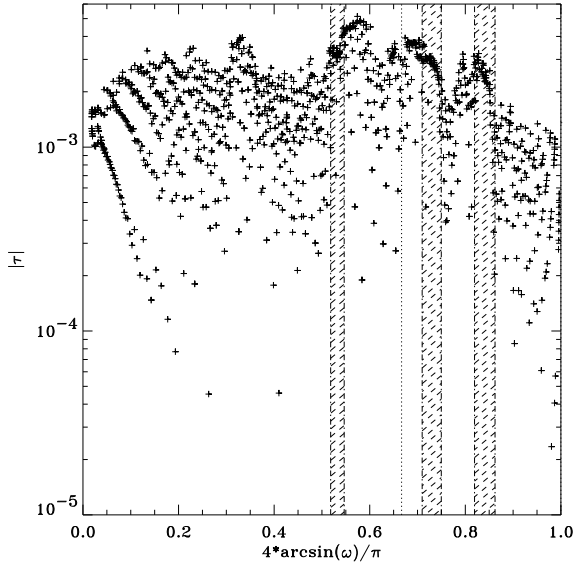,width=8cm}}
\caption{Distribution of the eigenvalues associated with least-damped
axisymmetric modes in the complex plane. Hatched frequency bands denoted
bands occupied by simple attractors; the dotted
line is for $\sin(\pi/6)$. The Ekman number is $10^{-8}$ and $\eta=0.35$.
We used a resolution of 700 spherical harmonics and 270 radial grid
points.}
\label{spectrum}
\end{figure}

This general evolution of the spectrum is well illustrated in
Fig.~\ref{spectrum}. Here, the least-damped eigenvalues have been
computed for $E=10^{-8}$. We clearly see frequency bands of attractors
avoided by weakly damped eigenvalues but see the gathering of these
eigenvalues around $\sin(\pi/4)$ and, but less conspicuously, around
$\sin(\pi/6)$.

To complete the picture, we need now mentioning that a few regular modes
survive among all these singularities; such modes are purely toroidal
modes or r-modes \cite{rieu00} which are non-axisymmetric. They avoid
the constraint of characteristics for their velocity field has no radial
component; this property makes their characteristics independent
of frequency (they are circles and vertical straight lines) and
authorizes smooth solution at zero viscosity. The associated eigenvalues
$\omega=1/(m+1), m\in\bbbn^*$ seem to be the only eigenvalues of the
Poincar\'e operator in a spherical shell.

Ending this letter, it is worth emphasizing the role of the geometrical
approach allowed by the dynamics of characteristics, for describing
the asymptotic properties of inertial modes; in the domain of
very low Ekman numbers ($10^{-10}\tv 10^{-20}$), typical of astrophysical
or geophysical fluids, these modes are out of reach numerically.

The foregoing presentation shows that inertial modes display a very rich
dynamical behavior which comes from the ill-posedness of the underlying
inviscid problem. Here we discussed the case of the spherical shell, but
our results are general and can be extended to any container; this
is important since natural containers are usually not perfect geometrical
objects. Hence, fortunately, a curve like Fig.~\ref{lyap} is structurally
stable (see our discussion relative to the core of the Earth in
\cite{rieu99}).

We note that the relevance of attractors has also been
shown experimentally in stratified fluids\cite{MBSL97}. Some configurations of
conducting fluids bathed by a magnetic field, obeying the PE, will also
display attractors\cite{Malkus67}. These properties are in fact very
general and extend to mixed-type PDE as illustrated by the case of
gravito-inertial modes\cite{DRV99}.  We think that similar results
should hold for systems which are solutions of PDE of hyperbolic or
mixed type meeting boundary conditions.  As an example, our results may
have applications in General Relativity and the problem of ``closed
timelike curves" (CTC), that is the problem of the existence of
physical systems which permit travels backward in time.  Such systems
like wormholes have been studied by various authors\cite{FMNEKTY90};
they set many problems among which that of causality. Such a problem is
also at the origin of the ill-posedness of the Poincar\'e problem and
we showed that it leads to many kinds of singularities.

We therefore see that inertial oscillations of a fluid inside a
container offers a paradigm which may guide our intuition for problems
in other fields of physics which are also ill-posed Cauchy problems.

We would like to thank Boris Dintrans and Leo Maas for very helpful
discussions. Part of the calculations have been carried out on the Cray
C98 of IDRIS at Orsay and on the CalMip machine of CICT in Toulouse
which are gratefully acknowledged.

\end{document}